# Hundredfold Enhancement of Light Emission via Defect Control in Monolayer Transition-Metal Dichalcogenides

D. Edelberg[1,†], D. Rhodes[2,†], A. Kerelsky[1], B. Kim[2], J. Wang[3], A. Zangiabadi[5], C. Kim[5], A. Abhinandan[2], J. Ardelean[1], M. Scully[4], D. Scullion[4], L. Embon[1], I. Zhang[1], R. Zu[5], Elton J. G. Santos[4], L. Balicas[5,6], Chris Marianetti[7], K. Barmak[2,#], X.-Y. Zhu[3,%], J. Hone[2,+], A. N. Pasupathy[1,*]

[1] Department of Physics, Columbia University, New York, NY 10027, USA
[2] Department of Mechanical Engineering, Columbia University, New York, NY 10027, USA
[3] Department of Chemistry, Columbia University, New York, NY 10027, USA
[4] School of Mathematics and Physics, Queen's University Belfast, BT7 1NN, UK
[5] National High Magnetic Field Laboratory, Florida State University, Tallahassee, Florida 32310, USA.
[6] Department of Physics, Florida State University, Tallahassee, Florida 32306, USA
[7] Department of Applied Physics and Applied Math, Columbia University, New York, NY 10027, USA

*Two dimensional (2D) transition-metal dichalcogenide (TMD) based semiconductors have generated intense recent interest due to their novel optical and electronic properties, and potential for applications. In this work, we characterize the atomic and electronic nature of intrinsic point defects found in single crystals of these materials synthesized by two different methods - chemical vapor transport and self-flux growth. Using a combination of scanning tunneling microscopy (STM) and scanning transmission electron microscopy (STEM), we show that the two major intrinsic defects in these materials are metal vacancies and chalcogen antisites. We show that by control of the synthetic conditions, we can reduce the defect concentration from above $10^{13}$ /cm$^2$ to below $10^{11}$ /cm$^2$. Because these point defects act as centers for non-radiative recombination of excitons, this improvement in material quality leads to a hundred-fold increase in the radiative recombination efficiency.*

The semiconducting transition metal dichalcogenides (TMDs) are a family of layered compounds with the formula $MX_2$ (M=[Mo, W] and X=[S, Se, Te]), which can be isolated in monolayer form and have promise in a wide range of applications in electronics and optoelectronics[1,2,3,4]. These materials host novel phenomena such as valley physics[5,6], interlayer tunneling[7,8], topological properties[9,10], and exciton superfluidity[5] which are of fundamental interest and may enable new device functionality. However, crystalline disorder obscures intrinsic phenomena and imposes an upper limit on achievable functionality[5,6,11,12]. In particular, point defects[13,14,15,16] strongly impact TMD monolayers: these defects cause carrier scattering and localization[17-19], act as centers for non-radiative recombination[20-23], and give rise to localized emission from excitonic traps[24,25]. Pioneering transport[7,26,27] and STEM studies[13,14,15] have explored the atomic nature and electronic impact of defects arising in monolayer TMDs, and indicate that the quality of these materials remains far behind the classic semiconducting materials such as Si and GaAs. Addressing the quality of these materials is urgently needed to advance their science and engineering applications.

In two-dimensional materials, disorder can arise from both intrinsic sources, such as point defects and grain boundaries in the crystal itself; and extrinsic sources arising from the environment, such as inhomogeneous strain, and charge traps / adsorbates in the substrate[14,28]. In the case of mechanically exfoliated graphene, the intrinsic defect density is extremely low ($10^9$-$10^{10}$/cm$^2$ [29,30]), and reducing extrinsic disorder by encapsulation in hexagonal boron nitride (hBN) has enabled spectacular advances in device performance[31-33]. However, other 2D materials do not necessarily possess graphene's ultrahigh purity. In the case of semiconducting TMDs, hBN encapsulation also results in improved performance[34,35], but the physical properties of these devices are still far from their theoretical limits, indicating that intrinsic disorder plays an important role. This is consistent with studies showing point defect densities exceeding $10^{12}$/cm$^2$ in commonly used TMD materials[36]. Therefore, continued progress in the field necessitates the characterization, quantification, and minimization of defects in TMD materials. Toward this end, a particular challenge is the diversity of material sources, which include natural or synthesized single crystals, and large-area films grown by chemical vapor deposition (CVD)[37,38], metal organic chemical vapor deposition[39], physical vapor deposition[40], and molecular beam epitaxy[41], each of which can give rise to a different density and type of defects. For instance, STEM imaging of $MoS_2$ reveals that CVD-grown films are dominated by S-vacancies, whereas defects in natural $MoS_2$ crystals are predominantly Mo vacancies[14].

In this study, we focus on the quality of synthesized TMD single crystals. Currently, single crystals remain the source of the highest-quality TMD monolayers[11,37,42,43], and do not suffer from grain boundaries and phase separation[16,44,45] observed in large-area films. While much initial work on TMDs has utilized naturally occurring minerals[7,46], laboratory-synthesized crystals provide a wider materials selection, and can offer a higher degree of quality control and reproducibility. Toward this end, a number of companies are currently supplying synthesized TMD crystals for laboratory use. However, synthesized TMD crystals have not been well characterized, and virtually no experimental work has examined the correlation between defect density and optoelectronic properties of monolayers derived from these bulk crystals. In this work, we use STM and STEM imaging to determine the type and density of intrinsic defects present in single crystals of $MoSe_2$ and $WSe_2$ synthesized by the chemical vapor transport (CVT) and flux growth techniques. Transition metal vacancies and anti-sites are found to be the most-common defect types, with Se vacancies being much rarer. Flux growth achieves 1-2 orders of magnitude lower defect density than CVT does, and defect densities lower than $10^{11}/cm^2$ from our flux-grown sample are by far the lowest reported for TMD materials. This improvement is reflected in reduction of band-edge disorder measured by scanning tunneling spectroscopy, and orders of magnitude increase in photoluminescence quantum yield.

The CVT technique utilizes a transport agent, usually a halogen, to transport starting materials from a hot region into a cooler growth region where they form crystals[47]. CVT provides large crystals in a relatively short growth time at moderate temperatures, and thus has become the prevalent technique for TMD synthesis. The self-flux method in which crystals are grown directly from the molten phase is an alternative method known to create higher-quality, albeit smaller, crystals[48]. In this work, we characterize three types of crystals: commercially obtained crystals grown by CVT and tested without further annealing (as-grown CVT or ag-CVT); crystals synthesized in our furnaces by CVT then annealed in a temperature gradient (treated CVT or t-CVT); and crystals grown by the self-flux method (flux). Details of the growth procedures, temperatures, and cooling rates are given in the methods section.

We first examine defects in bulk crystals through scanning tunneling microscopy (STM), which can provide defect lattice positions, local electronic structure, and defect density. To avoid surface contamination, crystals were cleaved *in situ* under UHV conditions. Examining a 25 nm square region of $MoSe_2$ (Figure 1a), we observe two defect types that can be initially identified by contrast as either "dark" or "bright". As discussed further below, these two predominant types of defects account for the vast (>99%) majority of defects imaged in both flux and CVT samples. Figures 1b and 1c show atomic resolution images of these defects. The bright defects, which we denote –X, are located on a selenium site (Fig. 1b). Since there is no missing atom associated with this defect, it is not a selenium vacancy[36,49], but is rather a substitutional impurity on the chalcogen site. Se vacancies can indeed be observed by STM but are roughly two orders of magnitude less common than the vacancies and substitutional impurities (see Supplementary Information section S1 for images of chalcogen vacancies). Further insight into this defect can be obtained by looking at the electronic structure using scanning tunneling spectroscopy (STS). Figure 1d shows a sequence of tunneling spectra measured at varying distances from a single -X defect, out to a distance of 3.5 nm. Directly over the defect, we measure a broad resonance pinned to the edge of the conduction band. This indicates that the –X defects behave as n-type dopants. STM itself cannot identify the chemical nature of defects. However, we have observed that the concentration of –X defects can vary widely between crystals depending on the growth method (see below), even when the same starting raw materials are used in the syntheses. This indicates that the defect is not associated with a foreign substituent, but is most likely an antisite defect, i.e, a Mo atom substituting for a Se atom as suggested by theoretical calculations (see below).

The dark defects observed in STM images, which we denote –M, are aligned with the Mo sites, which are located in the center of a triangle of selenium atoms (Fig. 1c). In contrast to the –X defects, the -M defects show a resonance near the edge of the valence band, as shown in Fig. 1e. This indicates that the –M defects are electron acceptors. Interestingly, whereas flux-grown $MoSe_2$ possesses both –M and –X defects, similarly grown $WSe_2$ displays only –M defects (see supplementary information S2 for atomic-resolution images of defects in $WSe_2$). This allows direct comparison of defects seen in STM and STEM. In order to perform STEM measurements, we exfoliate flakes of monolayer TMDs which are transferred onto holey carbon substrates as described in the Methods section. Shown in Fig. 1f is an STEM image that shows the only type of point defect observed in the flux-grown $WSe_2$ sample (see supplementary information S3 for addition STEM images). The bright atoms in this image are W due to its high atomic number, indicating that the defect is a missing W atom, i.e., a metal vacancy. This chemical assignment of the –M defect is consistent with the STS observation that they act as electron acceptors.

We quantify the defect density by large-area STM imaging. Figures 2a-c show topographic scans (0.5 µm x 0.5 µm) of the three $MoSe_2$ materials under study. In the ag-CVT sample (Fig. 2a), the defect density is high enough such that the individual point defects have overlapping electronic signatures, and therefore STM can only provide a lower bound on the defect density of $>10^{13}$ cm$^{-2}$ (1% of unit cells). This defect density is dramatically reduced, to $(2.5 \pm 1.5) \times 10^{12}$ cm$^{-2}$ (0.2%) in the t-CVT sample (Fig. 2b). The self-flux crystals display still lower defect density of $(1.7 \pm 0.5) \times 10^{11}$ cm$^{-2}$ (0.01%) (Fig. 2c). See supplementary information S4 for defect counting procedures and counts. As for $MoSe_2$, commercial ag-CVT $WSe_2$ exhibits a very high defect density and STM imaging can only provide a lower bound of $>10^{12}$ cm$^{-2}$ (0.1 %) (Fig. 2d). In the flux-grown $WSe_2$, the defect density is dramatically smaller, $(7.0 \pm 2.2) \times 10^{10}$ cm$^{-2}$ (0.006%) (Fig. 2e). This defect density is by far the lowest reported for any TMD semiconductor. We additionally sort defect counts by type for each growth method, a summary of these results for each growth are compared in Table 1.

Our STM and STS measurements can be compared to theoretical expectations of defect formation energies and electronic structure from density functional theory (DFT). Here we found that a metal vacancy requires an additional 5.22 eV per defect site for its formation (see Methods section for definitions of formation energies and chemical potentials). Using the Kröger-Vink (K-V) formalism to examine the chemistry of this defect gives a charge state of 4$^-$ when referenced to the neutral crystal causing this defect type to act as an electron acceptor (i.e., $\square_M^{4-} + 4h^\bullet$, where $\square_M^{4-}$ denotes the vacancy on the metal site for a compound with chemical formula $MX_2$ - this defect has a negative charge relative to the filled metal site in the neutral crystal reference state typically used in the K-V formalism, and $h^\bullet$ denotes the requisite holes in the valence band to maintain overall charge neutrality). DFT predictions for the local density of states match nicely with the observed STS, finding a shallow acceptor state (not pictured). The metal antisite defect has a formation energy of 4.81 eV. These defects exhibit a charge state of 6$^+$ when referenced to the neutral crystal making them electron donors, consistent with STM and STS (i.e., $M_X^{6+} + 6e'$ where $M_X^{6+}$ denotes the metal antisite and $e'$ represents the electrons required to retain charge neutrality). Apart from the two observed defect types in experiments, we also calculate the formation energy for chalcogen vacancies. We find this energy to be 1.81 eV, which is lower than that of a metal vacancy or antisite. The observed lack of these vacancies in spite of their lower formation energy suggests that kinetics plays a large factor in determining observed defect concentrations. Details of these calculations can be found in the methods and a table of calculated values can be seen in the supplemental information.

From the observed defect densities and binding energies, we can use semiconductor theory to calculate the chemical potential in various samples of $MoSe_2$ as a function of temperature[50]. The result of this calculation is shown in Figure 2f for the t-CVT and flux-grown samples. For the CVT material, we expect a relatively constant chemical potential due to large but compensated numbers of donor and acceptor defects. The flux-grown crystal, in contrast, has smaller defect density but a dominance of donors, resulting in a strong chemical potential shift towards the conduction band edge as a function of temperature. These considerations indicate that we should measure chemical potential shifts in our STS spectra between different samples. In order to measure this with high spatial resolution, we use STS spectroscopy to measure the local semiconducting gap at every pixel of a 256 x 256 px grid overlaid on a 0.5 x 0.5um$^2$ area. At each point, we extract a local value of the conduction and valence band edges from the local spectrum (see supplementary information S5 for a detailed description of the procedure). A sum of these two values gives the local value of the semiconducting bandgap.

Shown in Figure 3a is a color scale image of the measured bandgap variation in a t-CVT crystal. Color variations in this picture represent gap variations in the vicinity of defects. The average bandgap in this crystal is 860 meV, with defect-induced gap variations of order 50 meV. The bulk bandgap is in reasonable agreement with the theoretical gap of 840 meV [51]. The variation seen for the t-CVT crystal is to be contrasted with a similar gap variation image for the flux crystal shown in figure 3b. While the flux crystal also exhibits an average bandgap of 860meV, it displays much smaller variations in space due to the lower concentration of defects. To visualize these differences, histograms detailing the spread of gap sizes for t-CVT and flux are plotted in figure 3c and figure 3d respectively. The observed gap variation in the t-CVT ($3\sigma$) is 50 meV, while the flux crystal shows a gap variation of 20 meV. As part of our measurement, STS was able to extract the valence and conduction band edges separately. Therefore, we further the analysis by examining the impact defects have on the valence and conduction band edges. In figure 3f, 3f we plot the valence band onset distributions for t-CVT (centered at –370meV) and flux (centered at –760meV) respectively. From these plots we see that most of the gap variation arises from defect states on the valence edge, which we found earlier correspond to metal vacancies. A similar analysis of the conduction band edge is plotted for t-CVT in figure 3g (centered at 490meV) and flux in figure 3h (centered at 100meV). Here we see almost no variation, especially in the case of the flux crystal.

Our detailed gap maps can also be used to estimate the position of the chemical potential relative to the gap midpoint from $\mu = -\frac{<E_C> + <E_V>}{2}$. For the t-CVT sample, this gives a chemical potential 60 meV below the gap center, indicating slight p-type doping. The self-flux crystals have a chemical potential of 330 meV above the gap center, making them n-type. At 77K, we expect from figure 2f that the chemical potential should be 50 meV below the gap center for t-CVT and 320meV above the gap center for self-flux. This agreement with STS mapping over a 500x500nm$^2$ region indicates that we have properly accounted for all of the dopants in the semiconductor. Additionally, the observed behavior explains the commonly observed p-type ambipolar FET devices[52] that have been made from t-CVT crystals.

To connect the large disparity of crystal imperfections versus growth observed across different crystals in STM to the monolayer limit, we carry out photoluminescence (PL) measurements as a simple way to measure the radiative response versus its quality. We isolate single layers through mechanical exfoliation from bulk single crystals[37,42,43]. Monolayers were simultaneously exfoliated from ag-CVT, t-CVT and flux grown crystals and each sample was subsequently handled under identically conditions to eliminate extrinsic factors. Each monolayer was encapsulated in BN and placed on a passivated SiO$_2$ surface[34]. The resulting stacks were measured under the same conditions (laser excitation power, spot size, and acquisition time). The raw PL data for MoSe$_2$ at 4K are plotted on a log scale in figure 4a. While the peak position shifts very slightly as crystalline quality is enhanced, the linewidth (FWHM) shows an obvious decrease with improved crystal quality, from 4 meV in ag-CVT to 3 meV in t-CVT to 2 meV in self-flux sample. The decrease in FWHM is consistent with the improvement in homogeneity of monolayer sample as defect density decreases. A dramatic effect is seen in the total light emission intensity (proportional to the quantum yield), with the self-flux sample having a 10-fold increase in light intensity over the t-CVT monolayer, and a 100-fold increase over the ag-CVT monolayer. The large suppression of excitons with little change to peak position suggests that defects provide non-radiative pathways for the recombination of excitons, via exciton localization[53] or defect enhanced Auger processes[20,22,23].

To quantify the impact of defects on the excitonic properties, we extend the PL measurements described above to various temperatures on t-CVT and flux monolayers (ag-CVT is omitted due to a lack of intensity at high temperatures). We plot the data taken for both MoSe$_2$ (red) and WSe$_2$ (blue) on a log scale at 77K in figure 4b where t-CVT is dashed and flux is solid. Since the major observed difference from the crystals is the overall PL intensity, we plot the integrated PL intensity as seen for MoSe$_2$ in figure 4c. To model the shape of the integrated PL signal we must account for the unique band structure of monolayer TMDs. In the TMD materials, both the valence and conduction bands are spin split due to spin-orbit coupling. The magnitude of the splitting in the valence band is roughly an order of magnitude larger than the conduction-band splitting (~300 meV versus ~30 meV). Due to the spin splitting in the conduction band, one of the two transitions from the conduction band to the upper valence band (A exciton) is dark, while the other is optically bright. In MoSe$_2$ the lower of the two transitions is bright, while the situation is reversed in WSe$_2$. We fit the integrated PL intensity to an empirical Arrhenius equation, which accounts for both the spin split exciton[54], and non-radiative processes which average multiple recombination rates. The full analysis followed is described in the supplemental information S6. We find that the integrated PL emission can be described as follows for MoSe$_2$:

$$I_{Tot} = \frac{1}{1+C_1 * e^{-\frac{\Delta E_{NR}}{k_B T}} + C_2 * e^{-\frac{\Delta E_{Dark}}{k_B T}}}$$

Here the non-radiative term arises both from defects and phonon scattering at higher temperatures, and other term arises from thermal equilibrium between the dark and bright excitons. In figure 4c a sudden drop can be seen in the PL signal of the self-flux MoSe$_2$ monolayer above roughly 60K. This can be attributed to the Boltzmann distribution of electrons able to access the dark exciton state. As per our fit prescribed above we extract the dark exciton energy to be 40 meV above the bright exciton binding energy. This matches well with reports utilizing a backgate to directly measure the dark exciton at 30 meV above the bright one[55].

Like MoSe$_2$, the PL resonances of WSe$_2$ have roughly the same peak position across growth methods. The temperature dependent integrated intensity shows distinct behaviors for the flux versus CVT crystals as shown in figure 4d. The self-flux crystal shows an initial increase in the PL intensity with decreasing temperature down to about 150 K, below which the intensity drops sharply. The CVT crystal however shows a decreasing intensity with decreasing temperature starting from room temperature. This difference in the case of WSe$_2$ arises due to the fact that the lowest energy transition is dark[56]. This implies that at sufficiently low temperature only the dark state is populated, exponentially suppressing the PL intensity. At high enough temperature, on the other hand, the PL intensity is suppressed with increasing temperature due to phonon scattering. We can model both effects together using the same formula as MoSe$_2$ with a sign change:

$$I_{Tot} = \frac{1}{1+C_1 * e^{-\frac{\Delta E_{NR}}{k_B T}} + C_2 * e^{\frac{\Delta E_{Dark}}{k_B T}}}$$

This change of sign indicates that the dark exciton is now at a lower energy than the bright. Thus, PL shows a maximum at some intermediate temperature which is determined by the interplay between defect and phonon recombination versus dark exciton coupling. Using the fitting for the integrated intensity from the flux crystal, we extract a conduction-band splitting of 43 meV. Our measurement of this splitting is within experimental error of the value directly measured from magnetic field measurements (47 meV)[57,58]. We additionally find that CVT crystals exhibit stronger defect mediated coupling to the dark exciton state diminishing the PL signal out to much higher temperatures.

Our studies of defects in transition-metal dichalcogenides show definitively that there is a direct link between intrinsic point defect concentration in bulk crystals and the optoelectronic properties of exfoliated monolayers. The improvements in synthesis presented here have led to a lowering of the defect concentration by two orders of magnitude when compared to the current state of the art. Such improvements are a necessary step towards achieving many of the predicted optical phenomena that require high exciton concentration as well as transport phenomena that require long scattering times. We note in conclusion that while our synthetic achievements set a new benchmark for TMD semiconductors, the lowest bulk defect concentrations achieved here (~$10^{18}$ /cm$^3$) are still significantly higher than those achieved in the best III-V semiconductor films, indicating that there is still room for refining synthetic processes to achieve higher quality TMD materials.

## Methods

### Flux growth

MoSe$_2$ and WSe$_2$ crystals were synthesized by reacting Mo/W powders, 99.999%, with Se shot, 99.999%, typically in a ratio of 1:20. These materials were first loaded into a quartz ampoule. A piece of quartz wool is then pressed into a cylindrical shape and pushed into the quartz ampoule, approximately 1 cm above the raw elements. The ampoule was then evacuated and sealed at ~$10^{-3}$ Torr. For growth, the ampoule is heated to 1000 °C over 48 hours, held there for 3 days, then cooled at a rate of 1.5 °C down to 400 °C and subsequently flipped and centrifuged. Crystals are then harvested from the quartz wool filter and annealed at a temperature of 250 °C with the empty end of the quartz ampoule held approximately at 100 °C for 48 hours.

### CVT

MoSe$_2$ crystals were synthesized by reacting Mo powder, 99.999%, with Se shot, 99.999%, in stoichiometric proportions with iodine 99.999% as a transport agent. These materials were first loaded into a quartz ampoule 12 cm in length, 1 cm in diameter, then evacuated and sealed at ~$10^{-3}$ Torr. For growth, the ampoule is heated to 1000 °C over a period of 48 hours, held there for 1 week, then cooled for 3 days to 750 °C and subsequently quenched in air. Crystals are then harvested and rinsed in acetone and isopropanol to remove iodine residue, and left to dry.

### *Ab initio* density functional theory methods

First-principle calculations for defect formation energies were done using density functional theory (DFT) within the projected augmented wave method [59,60], as implemented in the VASP code [61,62]. The generalized gradient approximation [63] is employed to treat exchange and correlation in DFT. Projected augmented wave method (PAW) was used in the description of the bonding environment for W, Mo, and Se. The structures are fully relaxed until all interatomic forces are smaller than 0.02 eV/Å. The Brillouin zone was sampled with a 5×5×1 k-mesh under the Monkhorst-Pack scheme [64]. Plane-wave energy cut offs of 400 eV and 500 eV are used for structural relaxation and static runs, respectively. The defect formation energies are defined via $E_{form}=E_{defect}-(E_{pristine}+\Sigma n_i \mu_i)$, where $E_{defect}$ stands for the total energy of a defected monolayer, $E_{pristine}$ is the total energy of a pristine monolayer, $n_i$ is the number of removed (minus sign) or added (plus sign) species i and $\mu_i$ is the chemical potential of species i. The size of the supercell was determined by convergence tests, resulting in a 6×6×1 with a 15 Å vacuum space. The chemical potentials of each species are constrained by the relation [64], $\mu_{MoSe2}= \mu_{Mo}+2\mu_{Se}$, where $\mu_{Mo}$ and $\mu_{Se}$ are the chemical potentials for Mo and Se, respectively; $\mu_{MoSe2}$ is the total energy per formula unit of MoSe$_2$. We determine the range of chemical potentials with two extreme cases: Mo-rich environment and Se-rich environment. For the Mo-rich environment, $\mu_{Mo}$ is chosen to be the total energy per atom of Mo in the bcc structure. For the Se-rich environment, $\mu_{Se}$ is chosen to be the total energy per atom of Se in the trigonal phase. The defect formation energies as a function of $\mu_{Mo}$ are presented in the supplementary information S5. The values in the main text are extracted with the chemical potentials following $\mu_i=E_i+E_{bond}$, where $E_i$ = Total energy of bulk metal or crystal of chalcogenide and $E_{bond} = (E_{MX2}-E_M-2E_X)/3$ [65].

### Scanning Tunneling Microscopy and Spectroscopy (STM)

STM measurements were performed using a custom built, variable temperature, UHV STM system. Single crystals of MoSe$_2$ and WSe$_2$ were mounted onto metallic sample holders using a vacuum safe silver paste. Samples were then transferred into the STM chamber and cleaved in-situ, exposing a clean surface. A Pt-Ir STM tip was cleaned and calibrated against a gold (111) single crystal prior to the measurements. Measurements were collected at 82 K and 300 K.

### Optical measurements

Optical stacks of BN/TMD/BN were fabricated using the polypropylene carbonate (PPC) method as described in [66] and placed on passivated $SiO_2$[34]. $MoSe_2$ samples were measured using a closed-cycle He cryostat (Attocube Attodry 1100) and an excitation wavelength of 532 nm using a cw diode laser with an approximate power of 2.0 µW. For $WSe_2$, samples were loaded into a cryostat with a sapphire window which is combined with a homemade photoluminescence setup with an excitation wavelength of 532 nm using a cw diode laser and a power of 80 µW. For cooling, either helium-4 or liquid nitrogen were continuously flowed through the cryostat chamber, immersing the sample while temperature was modulated with a stage heater.

**Transmission Electron Microscope (TEM) sample preparation and Scanning Transmission Electron Microscope (STEM) Imaging**

For the preparation of electron transparent samples for point defect density measurements, the TMD crystals were mechanically exfoliated using Scotch™ tape. The exfoliated flakes were transferred onto oxidized silicon wafer substrates. The monolayer flakes of the exfoliated TMD crystals were identified using light optical microscopy. The monolayers were then transferred onto Quantifoil® holey carbon TEM grids using isopropyl alcohol as a medium. After the alcohol evaporated, the holey carbon grid was attached to the wafer and the monolayer. The wafer sections were then slowly immersed in 1M potassium hydroxide (KOH) solution to etch the very top surface of the oxide and release the TEM grid and the exfoliated crystals attached to it. Distilled water was used to dilute and wash away the KOH solution from the TEM grids. As a last step, TEM grids were immersed in warm (40 °C) acetone for 10 minutes to dissolve any residue that remained from the exfoliation and sample transfer.

The STEM imaging of TMD monolayers was carried out in an FEI Talos F200X instrument operated at 200 kV. A low beam current (~60 pA) was used to reduce the amount of beam damage. By using the smallest condenser aperture (50 µm) and beam size 9, the convergence angle of the probe was calculated to be approximately 10 mrad. Images were acquired at a series of times to determine the rate at which metal vacancies formed in the samples when irradiated by the electron beam. For $MoSe_2$, the density of metal vacancies was found to increase linearly with time allowing extrapolation to time zero to determine the initial density. For $WSe_2$, no new metal vacancies were found to be created up to 40 seconds of imaging. Thus, for these samples, images from multiple regions were collected using a 20 second acquisition time. To improve the contrast and reduce the noise in the images for quantification of point defect density, the Butterworth filter in Gatan Digital Micrograph and the Wiener deconvolution in MATLAB were used. In addition, the simulated structure of the monolayer using the CrystalMaker software was overlaid on the processed STEM images to aid in identification of the metal and the chalcogen positions.

# Acknowledgements

Yinan Dong, Bryan Medina and Steven French are acknowledged for their assistance in sample preparation. This work is supported by the NSF MRSEC program through Columbia in the Center for Precision Assembly of Superstratic and Superatomic Solids (DMR-1420634). Support for STM instrumentation is provided by the Air Force Office of Scientific Research (grant number FA9550-16-1-0601). KB and AZ acknowledge partial funding support from the Dean of the School of Engineering and Applied Science at Columbia University. DS acknowledges his EPSRC studentship. EJGS acknowledges the use of computational resources from the UK national high-performance computing service (ARCHER) for which access was obtained via the UKCP consortium (EPSRC grant ref EP/K013564/1); the UK Materials and Molecular Modeling Hub for access to THOMAS, which is partially funded by EPSRC (EP/P020194/1). EJGS also acknowledges the Queen's Fellow Award through the grant number M8407MPH, the Enabling Fund (A5047TSL), and the Department for the Economy (USI 097). L.B acknowledges the US Army Research Office MURI grant W911NF-11-1-0362 (Synthesis and Physical Characterization of Two-Dimensional Materials and Their Heterostructures) and the Office Naval Research DURIP Grant 11997003 (Stacking Under Inert Conditions). Preliminary growth and characterization single crystals were performed at the National High Magnetic Field Laboratory, which is supported by the NSF Cooperative Agreement DMR-1157490 and the State of Florida.

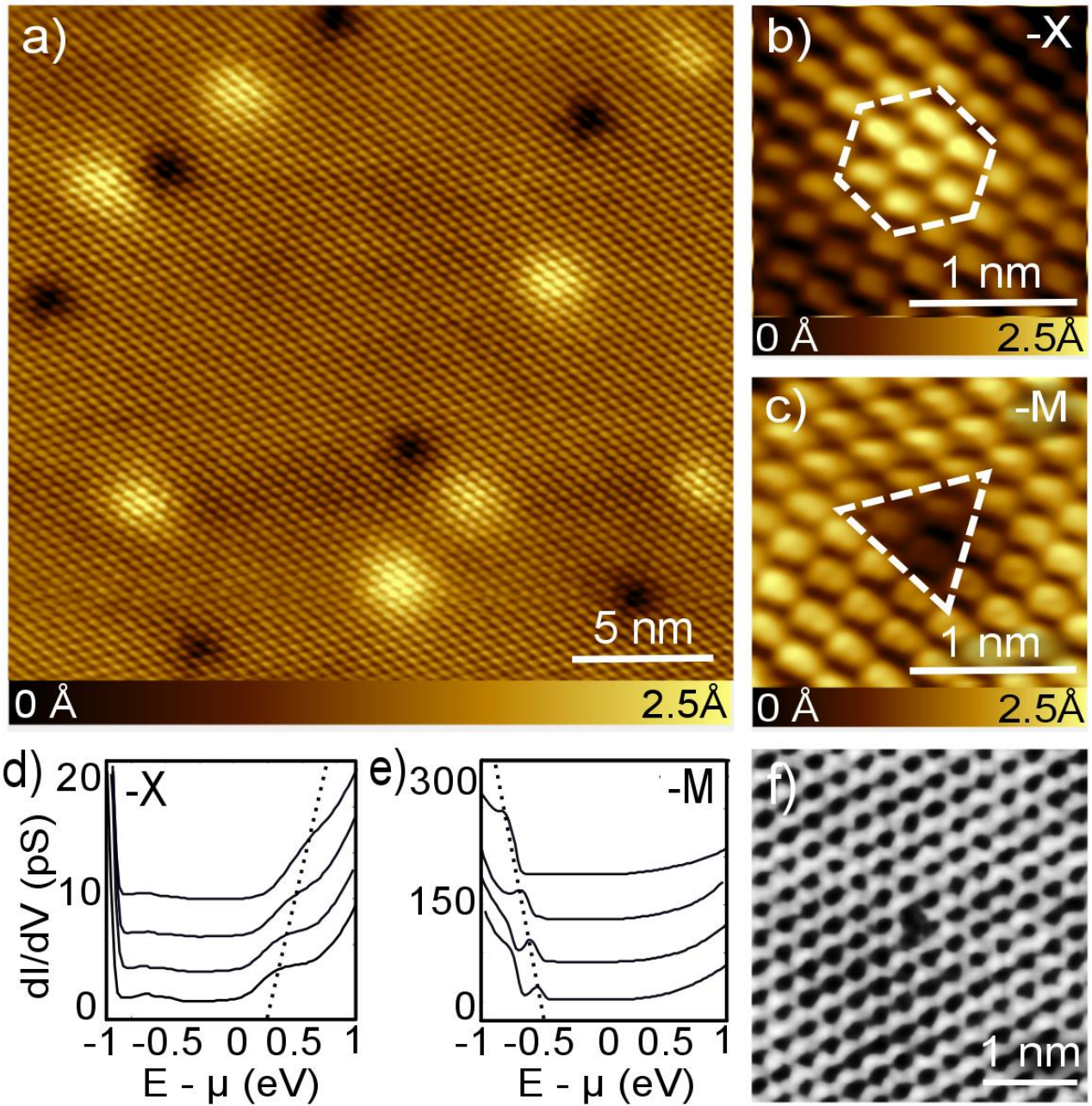

Figure 1 : Defect atomic and electronic structure
High resolution STM topographic images of MoSe$_2$ (V=1.25V  I=100 pA) showing a) a 25 nm area with two types of defects b) a single chalcogen antisite, -X type defect and c) a single metal vacancy, -M type defect. Differential conductance curves obtained at various distances from a d) chalcogen antisite, showing the presence of a donor state on the defect site, and e) single metal vacancy, showing the presence of an acceptor state on the defect site f) STEM imaging confirming that -M defects are metal vacancies

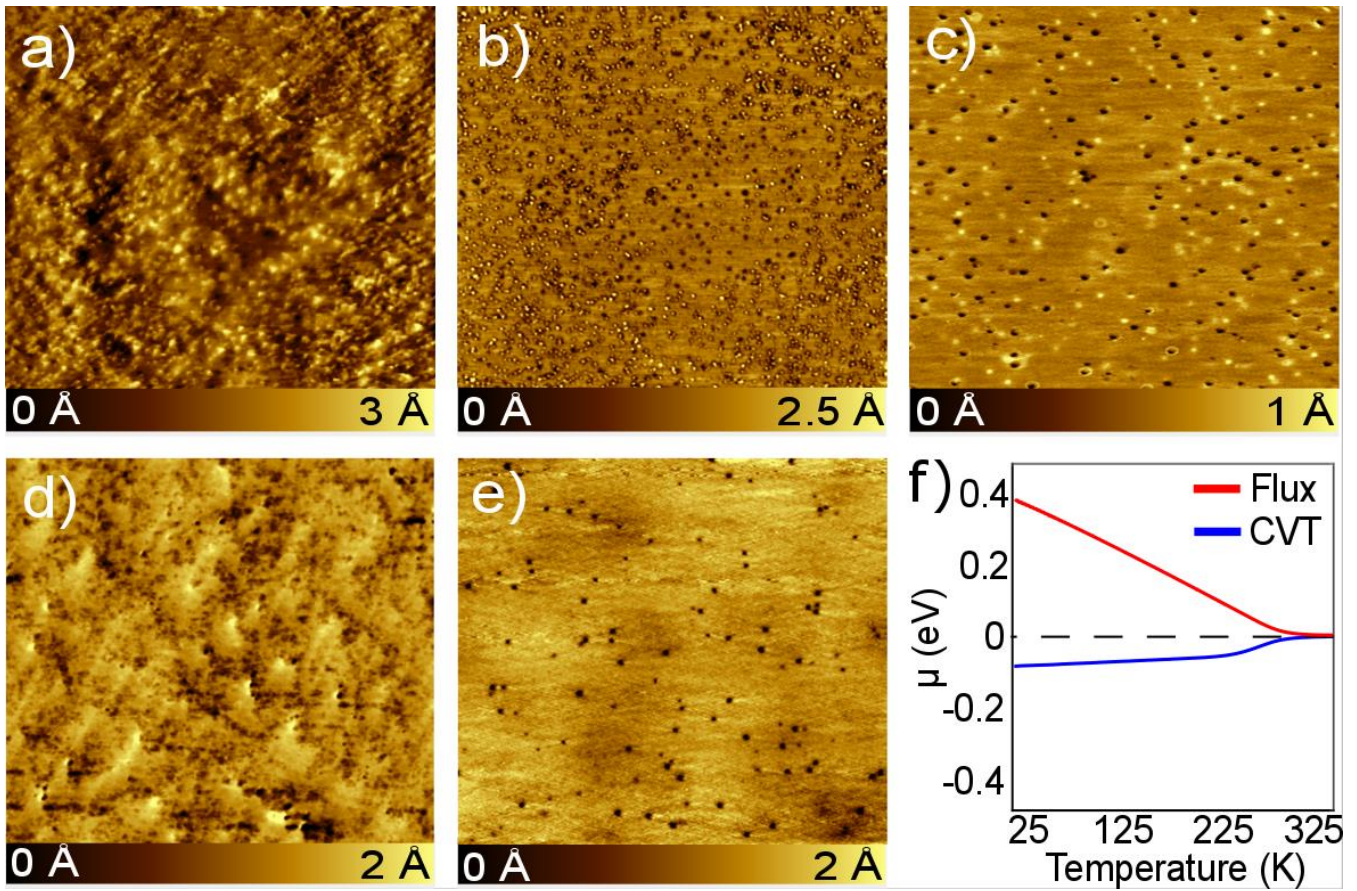

Figure 2 : Defect density versus growth method in MoSe$_2$ and WSe$_2$
STM topographic images of 500 x 500 nm$^2$ areas of an MoSe$_2$ crystal grown by a) commercial as-grown CVT b) post-annealed t-CVT, and c) self-flux. The imaging conditions for a-c) were a STM bias of 1.25 V and current of 100pA. STM topographic images of 500 x 500 nm$^2$ areas of WSe$_2$ crystals grown by d) commercial CVT (V=1V  I=100 pA)  and e) self-flux (V=0.75V  I=150 pA). f) Chemical potential calculated from defect densities for MoSe$_2$ on treated t-CVT and self flux crystals as a function of temperature.

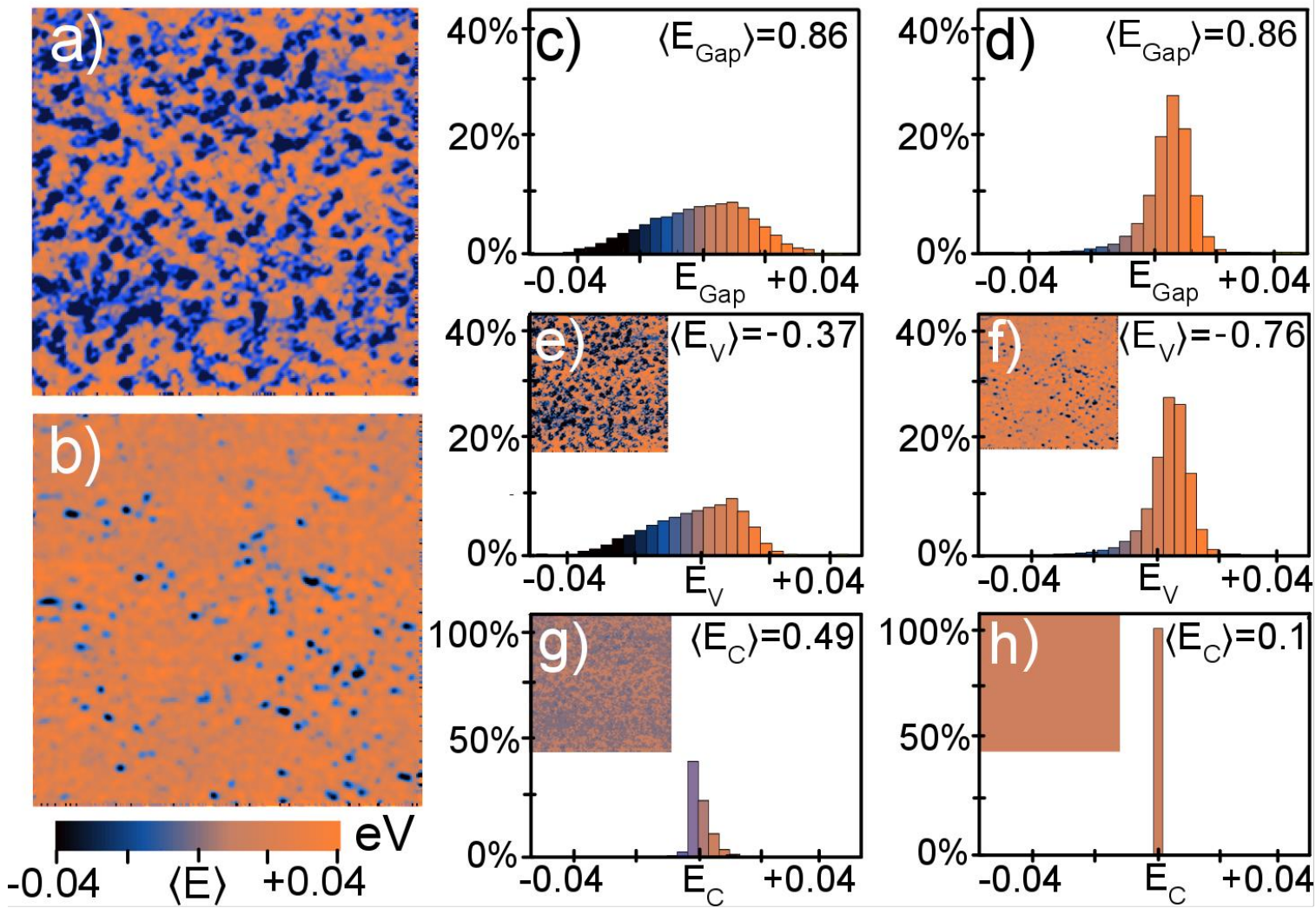

Figure 3: Spatially Resolved Bandgap by Scanning Tunneling Spectroscopy
Map of the semiconducting gap measured on a 256 x 256 pixel grid over a 500 x 500 nm$^2$ region plotted in the same color scale of a) post annealed t-CVT and b) self-flux grown MoSe$_2$ crystal at 82 K. c-d) Histograms of bandgap sizes based on a) and b) respectively. Histograms of e-f) valence band edge and g-h) conduction band edge from t-CVT and self-flux respectively. The insets to e-h) are measured images of the respective band edges in the same color scale.

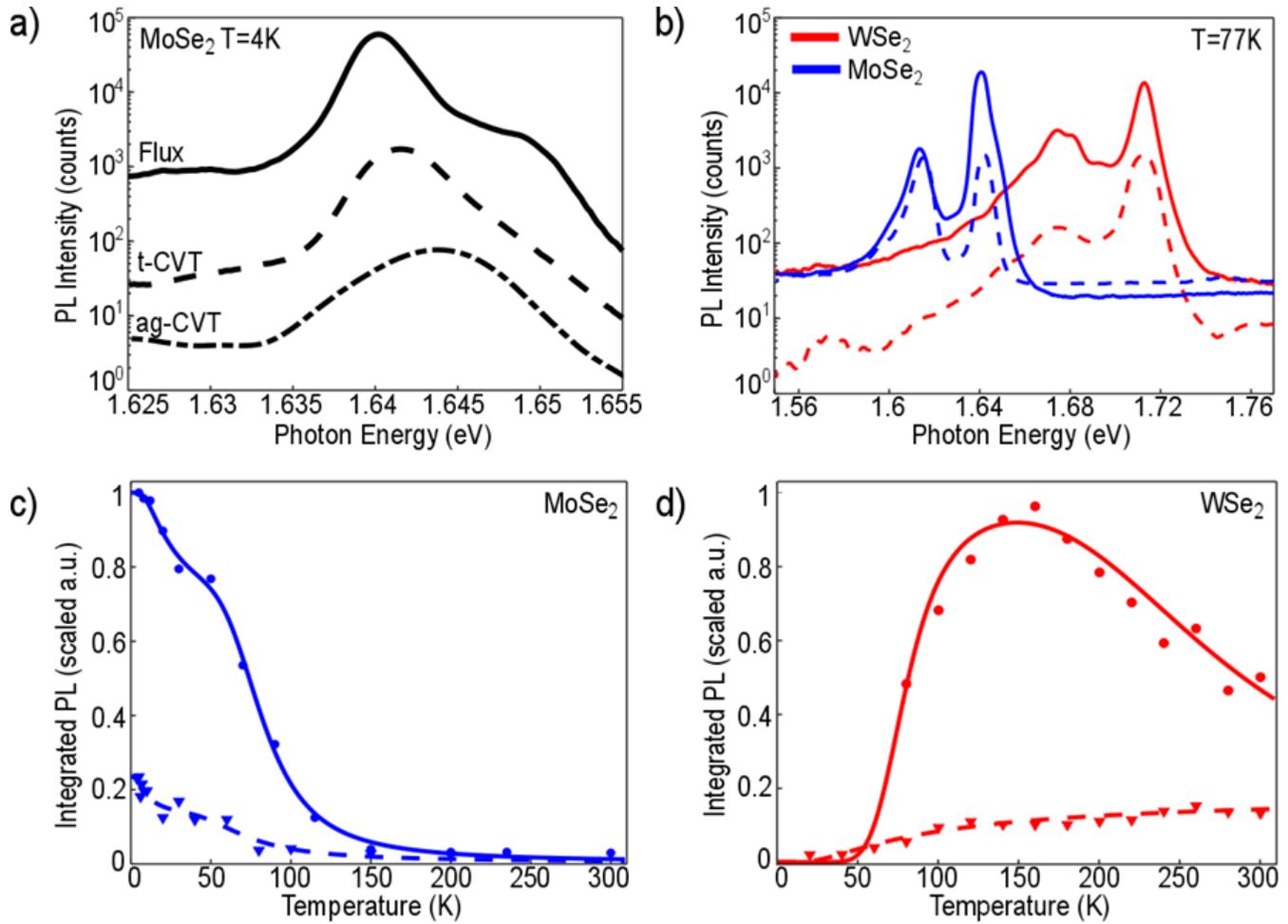

Figure 4: Temperature Dependent Photoluminescence of monolayer TMDs for different Growth Methods
a) PL for ag-CVT, t-CVT and self-flux grown monolayers of $MoSe_2$. All monolayers are encapsulated in h-BN and measured under identical conditions at 4K. The intensity of the PL from t-CVT and ag-CVT have been multiplied by 6.7 and 275 respectively in order to match the peak intensity of the flux-grown monolayer b) PL comparing t-CVT (dashed) and self-flux (solid) crystals of $MoSe_2$ (red) and $WSe_2$ (blue) c) Temperature dependence of the integrated A-exciton PL intensity in $MoSe_2$ for t-CVT and flux-grown samples. d) Temperature dependence of the integrated A-exciton PL intensity in $WSe_2$ for t-CVT and flux-grown samples. The drop in intensity at low temperature is due to the lowest excitonic transition being optically dark. The blue and red lines in b) and c) are fits to a modified Arrhenius form from which the conduction-band splitting can be extracted.

| Crystal Growth Method | Total Defect Count $Defects/cm^2$ | M- Site Defects $Defects/cm^2$ | X- Site Defects $Defects/cm^2$ |
|---|---|---|---|
| MoSe$_2$ ag-CVT | $> 10^{13}$ | $> 10^{13}$ | $> 10^{13}$ |
| MoSe$_2$ t-CVT | $(2.5 \pm 1.5) \times 10^{12}$ | $(1.3 \pm 0.8) \times 10^{12}$ | $(1.12 \pm 0.6) \times 10^{12}$ |
| MoSe$_2$ Self-flux | $(1.7 \pm 0.5) \times 10^{11}$ | $(7.0 \pm 0.2) \times 10^{10}$ | $(9.91 \pm 0.2) \times 10^{10}$ |
| WSe$_2$ ag-CVT | $> 10^{12}$ | $> 10^{12}$ | $> 10^{11}$ |
| WSe$_2$ Self-flux | $(7.0 \pm 2.2) \times 10^{10}$ | $(6.41 \pm 2.0) \times 10^{10}$ | $(5.83 \pm 1.8) \times 10^{9}$ |

Table 1: For each crystal studied, total defect density and distribution of defect types. The error bars are from sample to sample variation across different batches of crystals grown by the same growth method. Within a single growth batch, the statistical variations across the crystals are much smaller than the sample-to-sample variations across growth batches.

**References**

1. Xu, X., Yao, W., Xiao, D. & Heinz, T. F. Spin and pseudospins in layered transition metal dichalcogenides. *Nature Physics* **10**, 343 (2014).
2. Chen, X., Yan, T., Zhu, B., Yang, S. & Cui, X. Optical Control of Spin Polarization in Monolayer Transition Metal Dichalcogenides. *ACS nano* (2017).
3. Lee, J., Mak, K. F. & Shan, J. Electrical control of the valley Hall effect in bilayer MoS2 transistors. *Nat Nano* **11**, 421-425 (2016).
4. Mak, K. F. & Shan, J. Photonics and optoelectronics of 2D semiconductor transition metal dichalcogenides. *Nat Photon* **10**, 216-226 (2016).
5. Fogler, M., Butov, L. & Novoselov, K. High-temperature superfluidity with indirect excitons in van der Waals heterostructures. *arXiv preprint arXiv:1404.1418* (2014).
6. Sie, E. J., Frenzel, A. J., Lee, Y.-H., Kong, J. & Gedik, N. Intervalley biexcitons and many-body effects in monolayer ${\mathrm{MoS}}_{2}$. *Physical Review B* **92**, 125417 (2015).
7. RadisavljevicB, RadenovicA, BrivioJ, GiacomettiV & KisA. Single-layer MoS2 transistors. *Nat Nano* **6**, 147-150 (2011).
8. Ma, N. & Jena, D. Interband tunneling in two-dimensional crystal semiconductors. *Applied Physics Letters* **102**, 132102 (2013).
9. Soluyanov, A. A. *et al.* Type-II weyl semimetals. *Nature* **527**, 495-498 (2015).
10. Bruno, F. Y. *et al.* Observation of large topologically trivial Fermi arcs in the candidate type-II Weyl semimetal WT e 2. *Physical Review B* **94**, 121112 (2016).
11. Choi, W. *et al.* Recent development of two-dimensional transition metal dichalcogenides and their applications. *Materials Today* (2017).
12. Zeng, Q. & Liu, Z. Novel Optoelectronic Devices: Transition-Metal-Dichalcogenide-Based 2D Heterostructures. *Advanced Electronic Materials* (2018).
13. Lin, Z. *et al.* Defect engineering of two-dimensional transition metal dichalcogenides. *2D Materials* **3**, 022002 (2016).
14. Hong, J. *et al.* Exploring atomic defects in molybdenum disulphide monolayers. *Nat Commun* **6** (2015).
15. Zhou, W. *et al.* Intrinsic structural defects in monolayer molybdenum disulfide. *Nano letters* **13**, 2615-2622 (2013).
16. Yankowitz, M., McKenzie, D. & LeRoy, B. J. Local Spectroscopic Characterization of Spin and Layer Polarization in ${\mathrm{WSe}}_{2}$. *Physical Review Letters* **115**, 136803 (2015).
17. Walukiewicz, W. Carrier scattering by native defects in heavily doped semiconductors. *Physical Review B* **41**, 10218 (1990).
18. El-Mahalawy, S. & Evans, B. Temperature dependence of the electrical conductivity and hall coefficient in 2H-MoS2, MoSe2, WSe2, and MoTe2. *physica status solidi (b)* **79**, 713-722 (1977).
19. Fivaz, R. & Mooser, E. Mobility of charge carriers in semiconducting layer structures. *Physical Review* **163**, 743 (1967).
20. Wang, H., Zhang, C. & Rana, F. Ultrafast dynamics of defect-assisted electron–hole recombination in monolayer MoS2. *Nano letters* **15**, 339-345 (2014).
21. Palummo, M., Bernardi, M. & Grossman, J. C. Exciton radiative lifetimes in two-dimensional transition metal dichalcogenides. *Nano letters* **15**, 2794-2800 (2015).
22. Moody, G., Schaibley, J. & Xu, X. Exciton dynamics in monolayer transition metal dichalcogenides. *JOSA B* **33**, C39-C49 (2016).
23. Wang, H. *et al.* Fast exciton annihilation by capture of electrons or holes by defects via Auger scattering in monolayer metal dichalcogenides. *Physical Review B* **91**, 165411 (2015).
24. Bastard, G., Delalande, C., Meynadier, M., Frijlink, P. & Voos, M. Low-temperature exciton trapping on interface defects in semiconductor quantum wells. *Physical Review B* **29**, 7042 (1984).

25  Ky, N. H. & Reinhart, F. Amphoteric native defect reactions in Si-doped GaAs. *Journal of applied physics* **83**, 718-724 (1998).
26  Ovchinnikov, D., Allain, A., Huang, Y.-S., Dumcenco, D. & Kis, A. Electrical transport properties of single-layer WS2. *ACS nano* **8**, 8174-8181 (2014).
27  Baugher, B. W. H., Churchill, H. O. H., Yang, Y. & Jarillo-Herrero, P. Intrinsic Electronic Transport Properties of High-Quality Monolayer and Bilayer MoS2. *Nano Letters* **13**, 4212-4216 (2013).
28  Raja, A. *et al.* Coulomb engineering of the bandgap and excitons in two-dimensional materials. *Nature Communications* **8**, 15251 (2017).
29  Liu, L., Qing, M., Wang, Y. & Chen, S. Defects in graphene: generation, healing, and their effects on the properties of graphene: a review. *Journal of Materials Science & Technology* **31**, 599-606 (2015).
30  Zhong, J.-H. *et al.* Quantitative correlation between defect density and heterogeneous electron transfer rate of single layer graphene. *Journal of the American Chemical Society* **136**, 16609-16617 (2014).
31  Dean, C. R. *et al.* Boron nitride substrates for high-quality graphene electronics. *Nature nanotechnology* **5**, 722-726 (2010).
32  Mayorov, A. S. *et al.* Micrometer-scale ballistic transport in encapsulated graphene at room temperature. *Nano letters* **11**, 2396-2399 (2011).
33  Britnell, L. *et al.* Electron tunneling through ultrathin boron nitride crystalline barriers. *Nano letters* **12**, 1707-1710 (2012).
34  Ajayi, O. *et al.* Approaching the Intrinsic Photoluminescence Linewidth in Transition Metal Dichalcogenide Monolayers. *2D Materials* (2017).
35  Wang, J. I.-J. *et al.* Electronic transport of encapsulated graphene and WSe2 devices fabricated by pick-up of prepatterned hBN. *Nano letters* **15**, 1898-1903 (2015).
36  McDonnell, S., Addou, R., Buie, C., Wallace, R. M. & Hinkle, C. L. Defect-dominated doping and contact resistance in MoS2. *ACS nano* **8**, 2880-2888 (2014).
37  Lee, Y. H. *et al.* Synthesis of Large-Area MoS2 Atomic Layers with Chemical Vapor Deposition. *Advanced Materials* **24**, 2320-2325 (2012).
38  Chen, J. *et al.* Chemical Vapor Deposition of Large-size Monolayer MoSe2 Crystals on Molten Glass. *Journal of the American Chemical Society* (2017).
39  Eichfeld, S. M. *et al.* Highly scalable, atomically thin WSe2 grown via metal–organic chemical vapor deposition. *ACS nano* **9**, 2080-2087 (2015).
40  Muratore, C. *et al.* Continuous ultra-thin MoS2 films grown by low-temperature physical vapor deposition. *Applied Physics Letters* **104**, 261604 (2014).
41  Zhang, Y. *et al.* Direct observation of the transition from indirect to direct bandgap in atomically thin epitaxial MoSe2. *Nature nanotechnology* **9**, 111-115 (2014).
42  Van Der Zande, A. M. *et al.* Grains and grain boundaries in highly crystalline monolayer molybdenum disulphide. *Nature materials* **12**, 554-561 (2013).
43  Roy, A. *et al.* Structural and electrical properties of MoTe2 and MoSe2 grown by molecular beam epitaxy. *ACS applied materials & interfaces* **8**, 7396-7402 (2016).
44  Kerelsky, A. *et al.* Absence of a Band Gap at the Interface of a Metal and Highly Doped Monolayer MoS2. *Nano Letters* (2017).
45  Kang, K. *et al.* High-mobility three-atom-thick semiconducting films with wafer-scale homogeneity. *Nature* **520**, 656 (2015).
46  Manzeli, S., Ovchinnikov, D., Pasquier, D., Yazyev, O. V. & Kis, A. 2D transition metal dichalcogenides. **2**, 17033 (2017).
47  Ubaldini, A., Jacimovic, J., Ubrig, N. & Giannini, E. Chloride-Driven Chemical Vapor Transport Method for Crystal Growth of Transition Metal Dichalcogenides. *Crystal Growth & Design* **13**, 4453-4459 (2013).
48  Zhang, X. *et al.* Flux method growth of bulk MoS2 single crystals and their application as a saturable absorber. *CrystEngComm* **17**, 4026-4032 (2015).

49  Voiry, D., Mohite, A. & Chhowalla, M. Phase engineering of transition metal dichalcogenides. *Chemical Society Reviews* **44**, 2702-2712 (2015).
50  Jin, Z., Li, X., Mullen, J. T. & Kim, K. W. Intrinsic transport properties of electrons and holes in monolayer transition-metal dichalcogenides. *Physical Review B* **90**, 045422 (2014).
51  Tongay, S. *et al.* Thermally Driven Crossover from Indirect toward Direct Bandgap in 2D Semiconductors: MoSe2 versus MoS2. *Nano Letters* **12**, 5576-5580 (2012).
52  Pradhan, N. R. *et al.* Ambipolar molybdenum diselenide field-effect transistors: field-effect and hall mobilities. *ACS nano* **8**, 7923-7929 (2014).
53  Godde, T. *et al.* Exciton and trion dynamics in atomically thin MoSe 2 and WSe 2: Effect of localization. *Physical Review B* **94**, 165301 (2016).
54  Baranowski, M. *et al.* Dark excitons and the elusive valley polarization in transition metal dichalcogenides. *2D Materials* **4**, 025016 (2017).
55  Quereda, J., Ghiasi, T. S., van Zwol, F. A., van der Wal, C. H. & van Wees, B. J. Observation of bright and dark exciton transitions in monolayer MoSe 2 by photocurrent spectroscopy. *2D Materials* (2017).
56  Dery, H. & Song, Y. Polarization analysis of excitons in monolayer and bilayer transition-metal dichalcogenides. *Physical Review B* **92**, 125431 (2015).
57  Zhang, X.-X., You, Y., Zhao, S. Y. F. & Heinz, T. F. Experimental evidence for dark excitons in monolayer WSe 2. *Physical review letters* **115**, 257403 (2015).
58  Wang, G. *et al.* Magneto-optics in transition metal diselenide monolayers. *2D Materials* **2**, 034002 (2015).
59  Kresse, G. & Hafner, J. Ab initio molecular dynamics for open-shell transition metals. *Physical Review B* **48**, 13115 (1993).
60  Kresse, G. & Furthmüller, J. Efficient iterative schemes for ab initio total-energy calculations using a plane-wave basis set. *Physical review B* **54**, 11169 (1996).
61  Perdew, J. P., Burke, K. & Ernzerhof, M. Generalized gradient approximation made simple. *Physical review letters* **77**, 3865 (1996).
62  Blöchl, P. E. Projector augmented-wave method. *Physical review B* **50**, 17953 (1994).
63  Kresse, G. & Joubert, D. From ultrasoft pseudopotentials to the projector augmented-wave method. *Physical Review B* **59**, 1758 (1999).
64  Monkhorst, H. J. & Pack, J. D. Special points for Brillouin-zone integrations. *Physical review B* **13**, 5188 (1976).
65  Haldar, S., Vovusha, H., Yadav, M. K., Eriksson, O. & Sanyal, B. Systematic study of structural, electronic, and optical properties of atomic-scale defects in the two-dimensional transition metal dichalcogenides M X 2 (M= Mo, W; X= S, Se, Te). *Physical Review B* **92**, 235408 (2015).
66  Wang, L. *et al.* One-dimensional electrical contact to a two-dimensional material. *Science* **342**, 614-617 (2013).

† These authors contributed equally to this work

Correspondence to:
\* apn2108@columbia.edu
+ jh2228@columbia.edu
% xyzhu@columbia.edu
# kb2612@columbia.edu

# Supplementary Material:

## S1. Chalcogen vacancies in TMDs

In our STM imaging experiments, >99% of all of the point defects observed were either the metal vacancy (-M) or chalcogen antisite (-X). We have very occasionally observed chalcogen vacancies in the lattice as shown in figure S1. The chalcogen vacancy is observed as a depression in STM topography at all bias voltages, consistent with the missing atom on the surface. The depth of these chalcogen vacancies was measured to be 300pm, an order of magnitude deeper than other observed defect types. This defect does not have an observable in-gap state and does not affect the bandgap significantly. The density of chalcogen vacancies observed in our crystals was too low to estimate accurately the area density.

## S2. Atomic Resolution STM of $WSe_2$

Atomic resolution measurements were taken on $WSe_2$ as seen in figure S2a. These images appear with the opposite contrast when compared to those in the main text due to opposite choice of STM bias. As was the case of $MoSe_2$ two defect types were observed, one occurring on the metal site and one occurring on the chalcogen site. We again label the metal site defect as seen in figure S2b as a metal vacancy and the chalcogen site defect seen in figure S2c as a metal antisite. STM scans on both $WSe_2$ and $MoSe_2$ suggest that these defect types are common to the TMD family.

## S3. Additional STEM images

Flakes of TMD crystals were transferred onto holey carbon substrates as described in the methods section of the main text. Shown in Figure S3a is a large-area image of one of these flakes of flux grown $WSe_2$ having several different thicknesses of TMDs. A part of the flake (false colored in yellow) of monolayer thickness is examined for defects. Shown in figure S3b is a section of this flake with atomic resolution imaging. A single W vacancy is seen in this image. To prove that this defect was not caused by beam damage in these films, defect concentration was measured as function of image acquision time. Shown in figure S3c and 3d are roughly the same small area of pristine monolayer seen in figure 3b imaged for 20 seconds (figure S3c) and for 40 seconds total (figure S3d). Eight additional selenium vacancies are observed in figure S3d (circled in yellow) when compared to figure S3c, which has three vacancies (circled in black). From a systematic set of such imaging experiments, we have determined the intrinsic concentration of selenium defects is zero (as we see in figure3b) as well as the knock-off rate. We find that the beam damage is limited to knock-off at the selenium sites (which appear a 180 degree rotation to those at W sites), and that much longer times are needed to remove the W atoms. Therefore our count of W vacancies as described in the main text is unaffected by the STEM imaging conditions.

## S4. Defect counts and statistics from STM images

Chalcogen antisite (-X) and metal vacancy (-M) defects are easily distinguished from each other at an imaging bias of V=1.25 V and current of I=100 pA. Under these conditions, the –X defect appears as a roughly 10 pm bump in STM topography, as shown in figure S4a, while the –M defect appears as a roughly 10 pm depression (figure S4b). A large scale image such as the one shown in figure S4c can be processed by applying height thresholds to identify directly the –X (red) and –M (blue) defects, which can then be counted either manually or by edge detection algorithms.

## S5. Formation Energies

The defect formation energies as a function of μ_Mo are presented in the figure S5. A summary of theoretical results for charge state and formation energies as presented in the main text can be seen in Table S1. Refer to the methods section for the details of the first principles calculations.

## S6. Bandgap Extraction from STS

Bandgap values were found using an automated extraction method for each pixel of our spatially resolved spectroscopy maps. Since the measured dI/dV signal is a convolution of the density of states with the Fermi-Dirac distribution of thermally activated carriers, we use a linear extrapolation to extract the conduction and valence band edges as shown in figure S6a. This figure illustrates the procedure at a location far away from defects. A similar process is employed when spectra are taken over a defect as shown in figure S6b. The band edge is still extracted by linear extrapolation from higher energies as in the defect free case. To extract the energy of the defect state an additional resonance (assumed to be Gaussian in shape) is added near the band edge.

## S7. Modeling temperature-dependent photoluminescence

To model the integrated photoluminescence a modified Arrhenius like formula was used. This formula accounts for the presence of dark excitons as well as other non-radiative factors. To accurately represent this PL signature we first look at the band structure of a single layer. Shown in figure S7a is a simplified band structure of monolayer MoX$_2$ compounds. Due to orbital hybridization of the $d_{x^2-y^2}$ and $d_{xy}$ orbitals a spin splitting of (100 – 500 meV) occurs in the valence band. A smaller splitting of (10-100 meV) exists in the conduction band due to strong spin orbit coupling (SOC) of the transition metal orbitals. In figure S7b a three dimensional rendering of the A exciton bands can be seen. The two lowest energy excitons therefore corresponds to a hole in the top of the valence band and an electron in one of the two spin split conduction bands. The lowest energy exciton is bright while the other is dark in MoX$_2$ due to spin selection rules (figure S7c). The SOC splitting of the conduction band is opposite in the case of WX$_2$ causing the lowest energy exciton to be dark. For every photon that is absorbed there are three recombination channels [1] - radiative recombination via the bright A exciton, non-radiative recombination from the dark A exciton and defect/phonon assisted recombination. We can write the rates of each of these three processes (assuming them to be independent) in terms of their energies and rate constants [2]:

$$W_{PL} = C_{Bright} * e^{-\frac{E_{Bright}}{k_B T}}, W_{Dark} = C_{Dark} * e^{-\frac{E_{Dark}}{k_B T}}, W_{NR} = C_{NR} * e^{-\frac{E_{NR}}{k_B T}}$$

We can put these together to extract the temperature dependent integrated PL from the ratio of incoming to outgoing photons

$$MoX_2: \frac{I_{PL}}{I_{Tot}} = \frac{C_{Bright} * e^{-\frac{E_{Bright}}{k_B T}}}{C_{Dark} * e^{-\frac{E_{Dark}}{k_B T}} + C_{NR} * e^{-\frac{E_{NR}}{k_B T}} + C_{Bright} * e^{-\frac{E_{Bright}}{k_B T}}} = \frac{1}{C_1 * e^{-\frac{\Delta E_{SOC}}{k_B T}} + C_2 * e^{-\frac{\Delta E_{NR}}{k_B T}} + 1}$$

A similar analysis can be performed for WSe$_2$ where the only difference is the sign on the SOC term.

$$WX_2: \frac{I_{PL}}{I_{Tot}} = \frac{1}{C_1 * e^{\frac{\Delta E_{SOC}}{k_B T}} + C_2 * e^{-\frac{\Delta E_{NR}}{k_B T}} + 1}$$

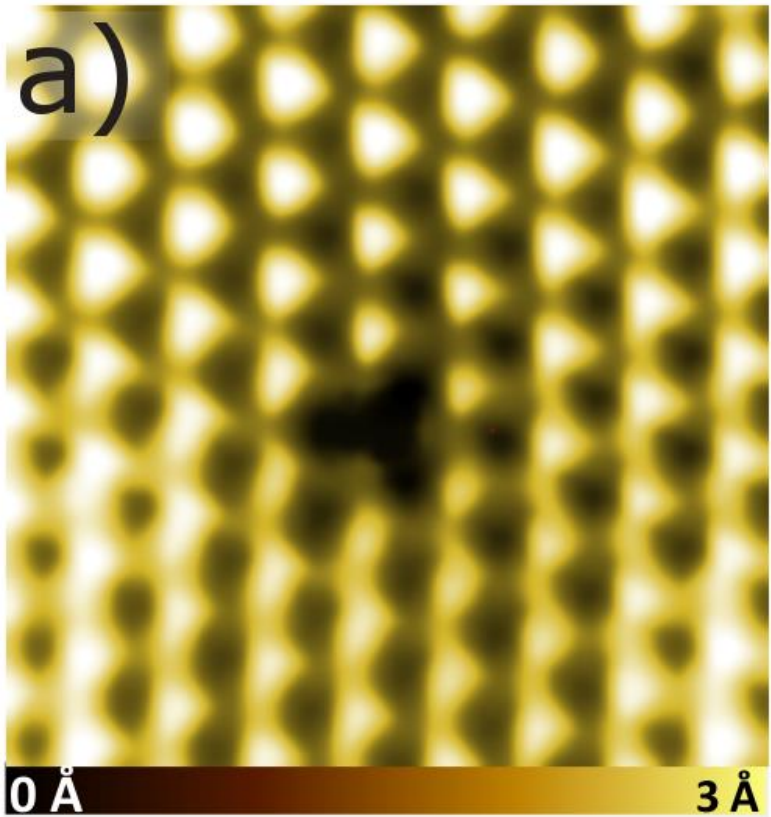

Figure S1: Chalcogen vacancies in STM

a) STM topograph of a single Se vacancy in $MoSe_2$ (V=1.4 V, I=160 pA). The vacancy shows up as a depression at all bias voltages.

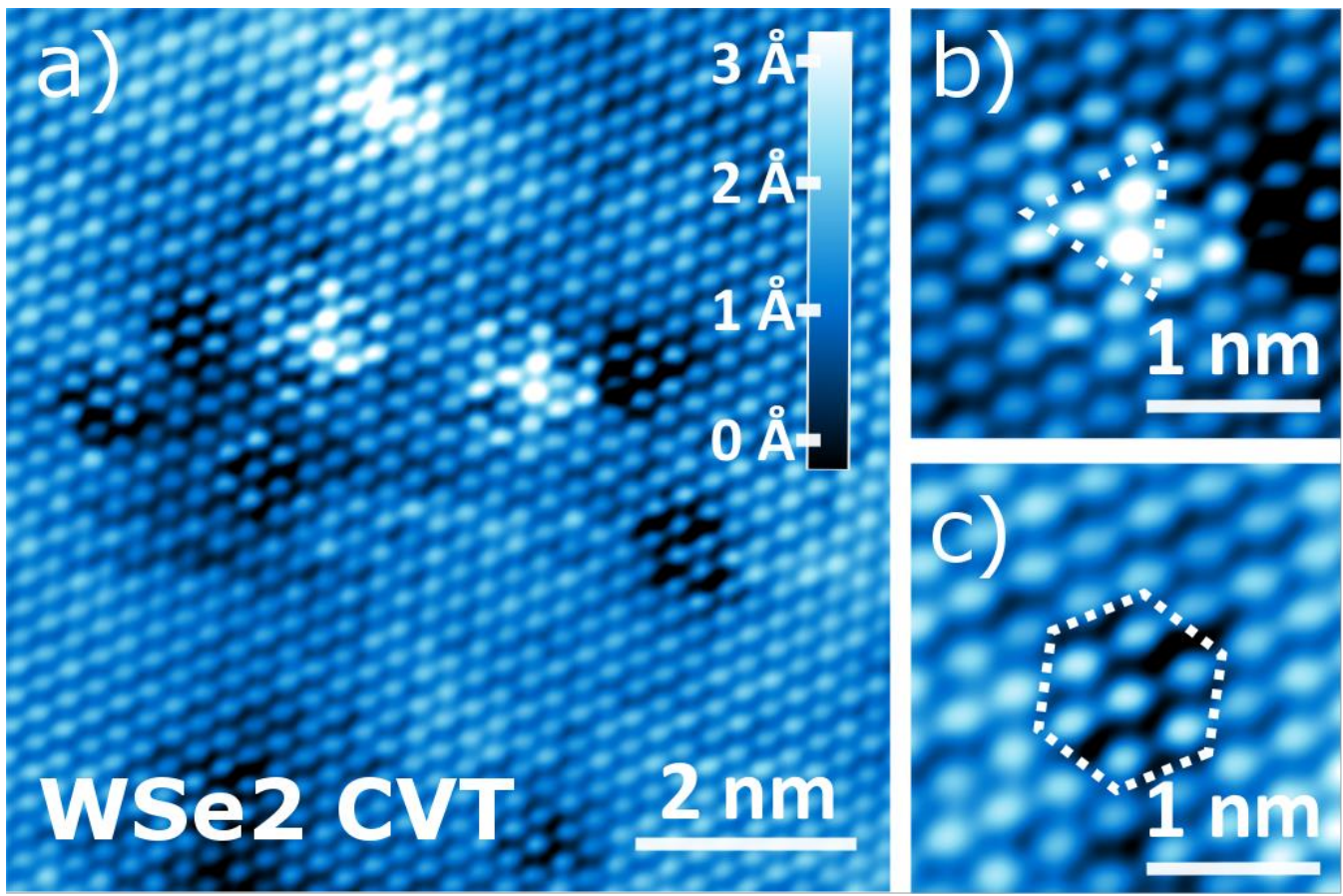

Figure S2: Atomic Resolution of Defects in WSe$_2$
a) Atomic resolution of an 8 nm$^2$ area of WSe$_2$ (imaged at -100 pA, -1V) where multiple defects can be seen b) Zoom in on a tungsten metal site defect located between chalcogen atoms c) Zoom in on a selenium antisite defect

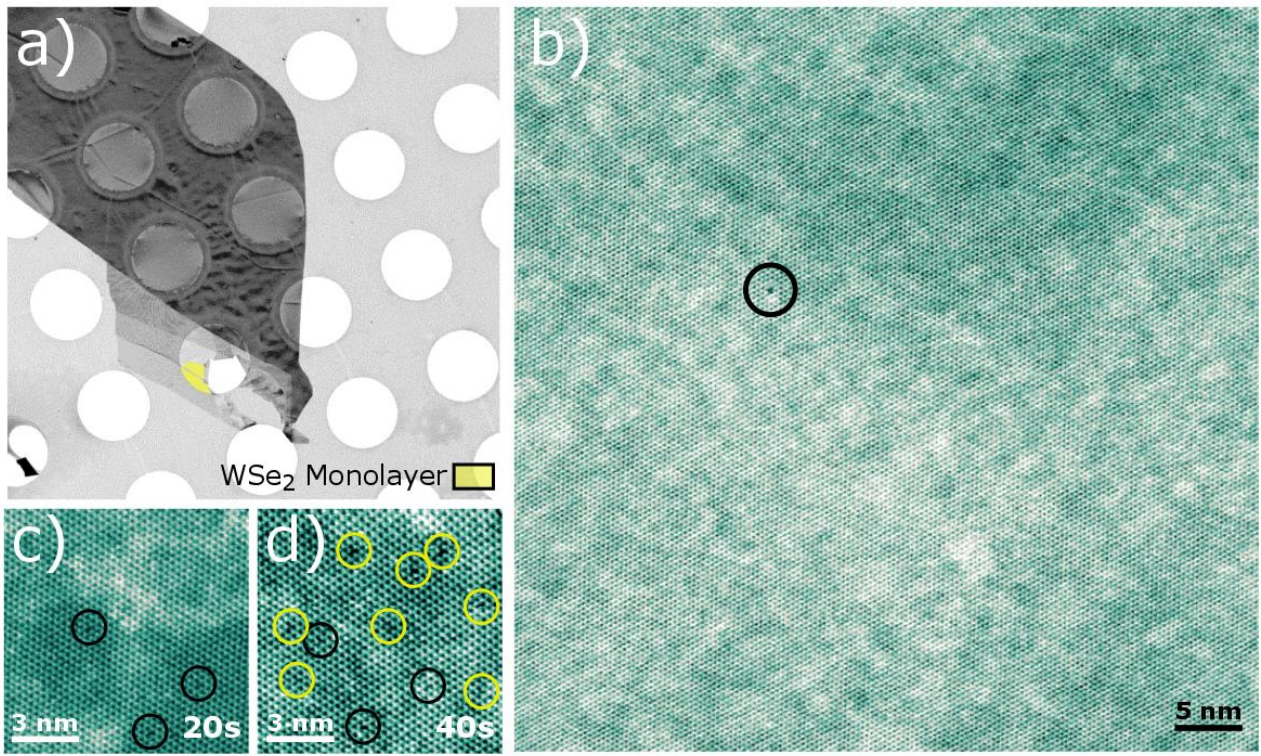

Figure S3: Scanning transmission electron microscopy imaging

a) Large-scale image of a WSe$_2$ flake showing monolayer and thicker layers on a holey carbon grid b) Large scale STEM image showing a single W vacancy in the field of view. c,d) Irradiation damage of the monolayer. Shown in c is a STEM image obtained with 20 second acquisition time on a small area of b showing three additional selenium vacancies in the field of view. d is the same area of the sample after 40 seconds of imaging, showing several additional beam-induced selenium vacancies. From systematic measurements of beam-induced damage as a function of acquisition time, the effect of beam damage can be accurately estimated and eliminated, giving a true measure of the intrinsic defect density in the crystal.

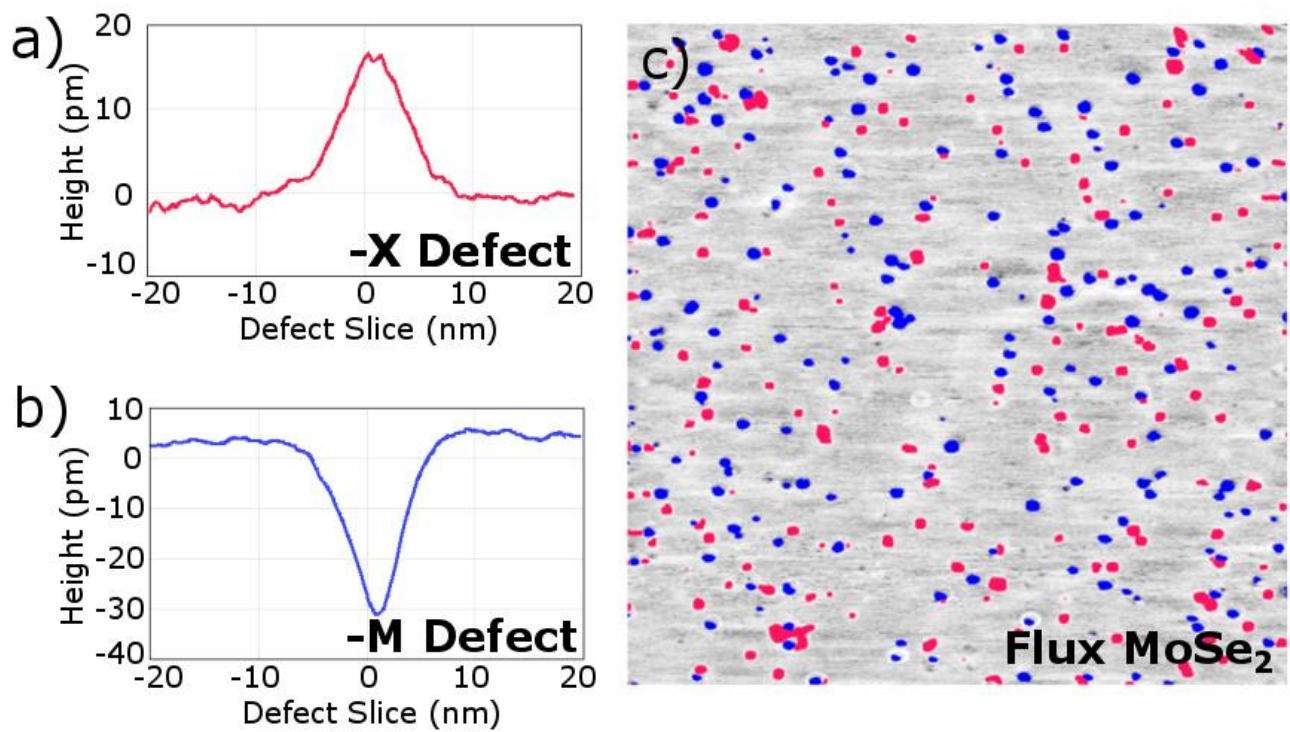

Figure S4: Defect identification and count

a,b) Identification of defects in MoSe$_2$. Chalcogen antisites (-X) and metal vacancies (-M) are clearly visible in STM topographies at V=1.25 V I=100 pA as a topographic bump and depression respectively. c) Defects can be counted by simply thresholding the STM topographic image as shown and counting the total number of continuous dark and bright patches.

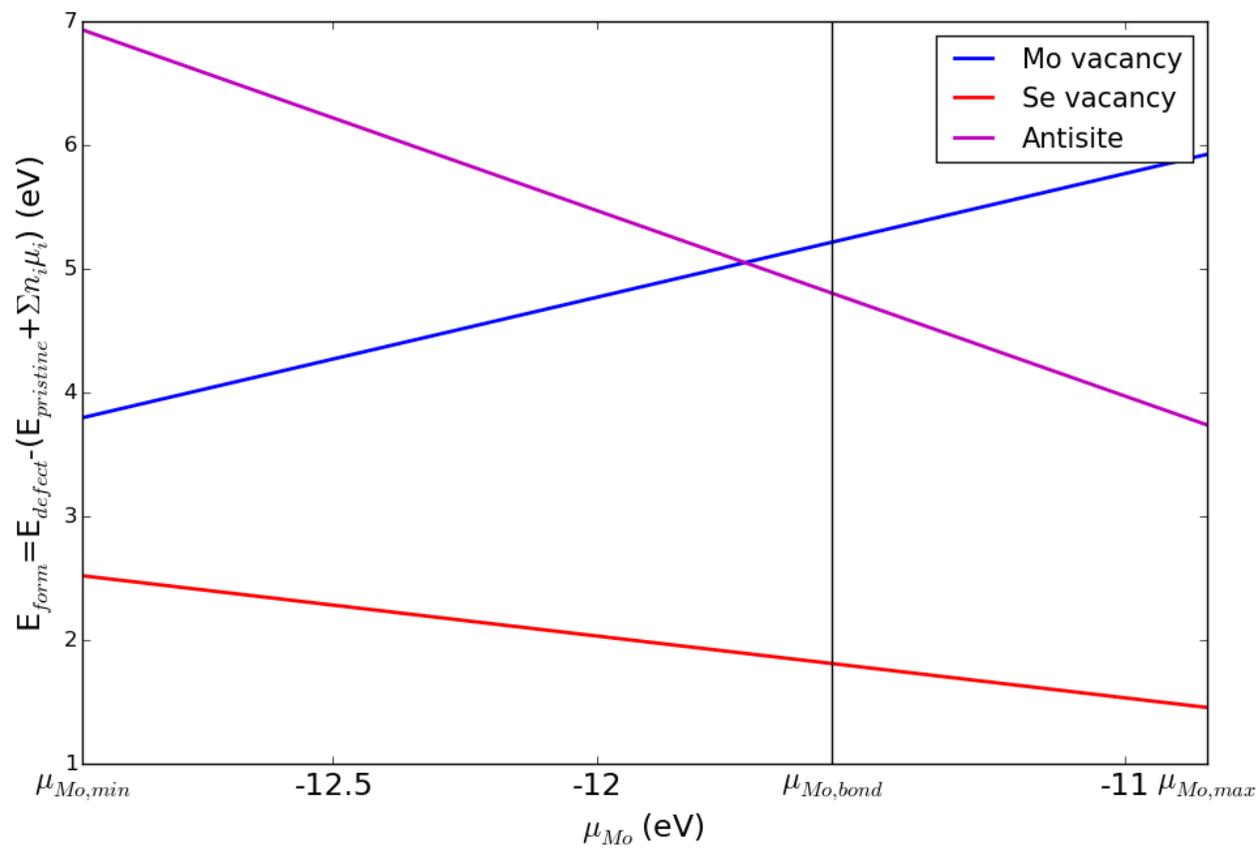

Figure S5: Defect formation energies

Calculated defect formation energies from DFT as a function of the chemical potential of Mo.

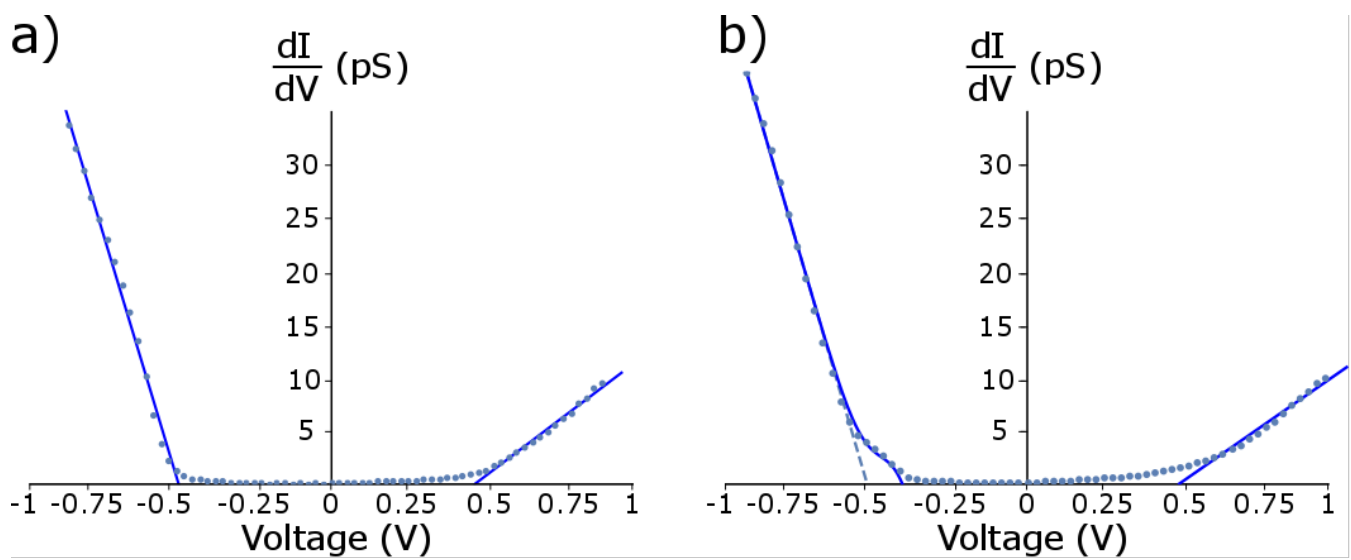

Figure S6: Band Edge Extrapolation and Defect Peak Identification
a) Automated linear extrapolation of valence and conduction band edge, for a spectrum taken at a defect free location b) Automated extrapolation of valence and conduction band edges, for a spectrum taken over a defect. A defect state is seen near the valence band edge, whose position can be determined by fitting an additional Gaussian peak to the linearly extrapolated band edge.

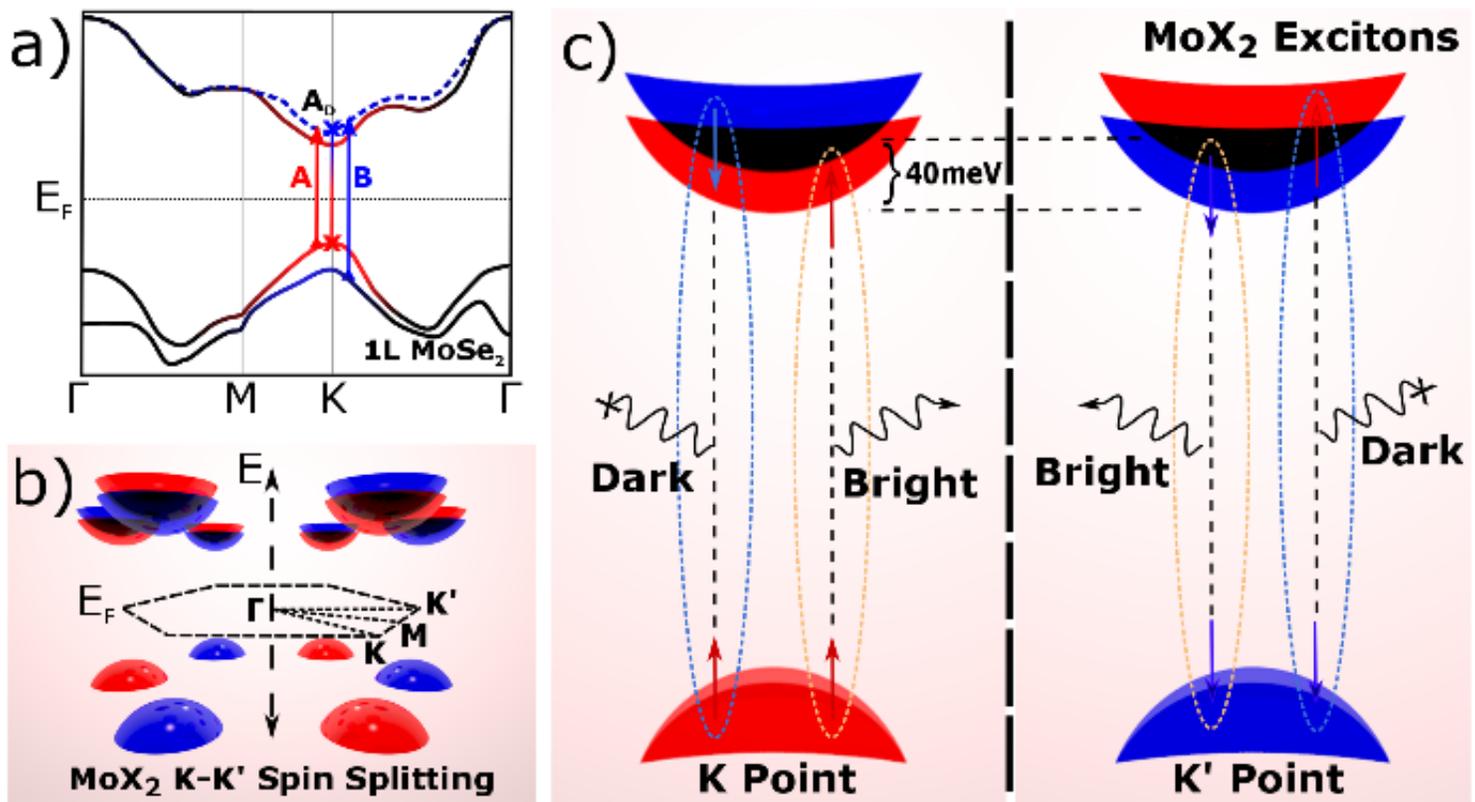

Figure S7: Exciton Dynamics in MoX$_2$ TMDs
a) Simplified band structure of 1L MoSe$_2$ showing A$_{Bright}$, A$_{Dark}$, and B excitons b) Three dimensional rendering of the band structure showing only the band edges at K and K' c) Band structure at the K and K' point showing the dark and bright exciton as determined by spin selection rules of the conduction electrons

| Defect Type | Kröger-Vink | Formation Energy | Charge State |
|---|---|---|---|
| Mo Vacancy | $\square_M^{4-}$ | 6.68 eV | $4^-$, Acceptor |
| Mo Antisite | $M_X^{6+}$ | 5.04 eV | $6^+$, Donor |
| Se Vacancy | $\square_X^{2+}$ | 1.72 eV | $2^+$, Donor |

Table S1: A summary of theoretical results presented for each defect type.

# References

1    Fang, Y. *et al.* Investigation of temperature-dependent photoluminescence in multi-quantum wells. *Scientific reports* **5** (2015).
2    Baranowski, M. *et al.* Dark excitons and the elusive valley polarization in transition metal dichalcogenides. *2D Materials* **4**, 025016 (2017).

† These authors contributed equally to this work

Correspondence to:
\* apn2108@columbia.edu
\+ jh2228@columbia.edu
% xyzhu@columbia.edu
\# kb2612@columbia.edu